\documentclass[12pt,preprint]{aastex}
\usepackage{apjfonts, emulateapj5, natbib}
\input psfig.tex

\def\aa{{A\&A}}

\def\aj{{AJ}}

\def\annrev{{ARA\&A}}
\def\apj{{ApJ}}
\def\apjs{{ApJS}}
\def\asec{$^{\prime\prime}$}

\def\cc{cm$^{-3}$}

\def\cc{cm$^{-3}$}

\def\flamb{erg s$^{-1}$ cm$^{-2}$ \AA$^{-1}$}

\def\ha{H$\alpha$}
\def\hb{H$\beta$}

\def\kms{km s$^{-1}$}
\def\lamb{$\lambda$}
\def\lax{{$\mathrel{\hbox{\rlap{\hbox{\lower4pt\hbox{$\sim$}}}\hbox{$<$}}}$}}
\def\gax{{$\mathrel{\hbox{\rlap{\hbox{\lower4pt\hbox{$\sim$}}}\hbox{$>$}}}$}}
\def\simlt{\lower.5ex\hbox{$\; \buildrel < \over \sim \;$}}
\def\simgt{\lower.5ex\hbox{$\; \buildrel > \over \sim \;$}}
\def\lum{erg s$^{-1}$}
\def\mbh{{$M_{\rm BH}$}}

\def\mnras{{MNRAS}}

\def\pasp{{PASP}}

\def\percm2{cm$^{-2}$}

\def\solmass{$M_\odot$}

\def\oiii{[\ion{O}{3}]}

\def\nii{[\ion{N}{2}]}

\def\mbh{{$M_{\rm BH}$}}

\slugcomment{To appear in {\it The Astrophysical Journal Letters.}}
\lefthead{Ho et al.}
\righthead{}

\begin{document}

\title{The Low-mass, Highly Accreting Black Hole Associated with the Active Galactic Nucleus 2XMM~J123103.2+110648}

\author{Luis C. Ho\altaffilmark{1}, Minjin Kim\altaffilmark{1,2}, and Yuichi 
Terashima\altaffilmark{3}}

\altaffiltext{1}{The Observatories of the Carnegie Institution for Science, 
813 Santa Barbara Street, Pasadena, CA 91101, USA}

\altaffiltext{2}{Korea Astronomy and Space Science Institute, Daejeon 305-348, 
Republic of Korea}

\altaffiltext{3}{Department of Physics, Ehime University, Matsuyama, Ehime 
790-8577, Japan}

\begin{abstract}
Optical spectra and images taken with the Baade 6.5 meter Magellan telescope 
confirm that 2XMM~J123103.2+110648, a highly variable X-ray source with an 
unusually soft spectrum, is indeed associated with a type 2 (narrow-line) 
active nucleus at a redshift of $z=0.11871$.  The absence of broad \ha\ or 
\hb\ emission in an otherwise X-ray unabsorbed source suggests that it 
intrinsically lacks a broad-line region.  If, as in other active galaxies, the 
ionized gas and stars in J1231+1106 are in approximate virial equilibrium, and 
the $M_{\rm BH}-\sigma_*$ relation holds, the exceptionally small velocity 
dispersion of $\sigma = 33.5$ \kms\ for \oiii\ \lamb 5007 implies that \mbh\ 
$\approx\, 10^5$ \solmass, among the lowest ever detected.  Such a low black 
hole mass is consistent with the general characteristics of the host, a small, 
low-luminosity, low-mass disk galaxy.  We estimate the Eddington ratio of the 
black hole to be \gax 0.5, in good agreement with expectations based on the 
X-ray properties of the source.   
\end{abstract}


\section{Introduction}

Active galactic nuclei (AGNs) in low-mass, late-type galaxies provide important
insights into the demographics of nuclear black holes (BHs) on the bottom end 
of the BH mass function.  With typical masses \mbh\ \lax\ $10^6$ \solmass, 
this class of AGNs not only helps to illuminate accretion physics in a poorly 
explored regime of parameter space, but it also serves as local analogs for 
the seeds of supermassive BHs in quasars.  To date the bulk of the known 
low-mass BHs have been discovered in the optical (Filippenko \& Ho 2003; Barth 
et al. 2004, 2008; Greene \& Ho 2004, 2007c; Ai et al. 2011), supplemented 
by a handful of cases found through mid-infrared spectroscopy (e.g., Satyapal 
et al. 2009).  X-ray observations have additionally revealed a potentially 
important population of much more numerous low-mass BHs accreting at low rates 
in nearby late-type spiral galaxies (Desroches \& Ho 2009; Araya Salvo et al. 
2012).

X-ray variability---a near-universal attribute of AGNs---presents another 
potentially effective strategy to search for low-mass BHs.  As both the 
amplitude and timescale of AGN variability depend on mass (e.g.,
Papadakis 2004), lower mass objects are expected to vary 
more strongly and more rapidly.  Previous X-ray variability studies indeed 
support this trend (NGC~4395: Vaughan et al. 2005; POX~52: 
Thornton et al. 2008; SDSS-selected objects: Minniuti et al. 2009).  Recently, 
Kamizasa et al. (2012) utilized the {\it XMM-Newton}\ serendipitous source 
catalog 
to identify 15 variable X-ray sources that are 
good AGN candidates.  From the strength of the X-ray variability, they predict 
that these AGNs have \mbh\ $\approx\, (1-7) \times 10^6$ \solmass.  

Terashima et al. (2012) highlight the most extreme case from Kamizasa et al.'s
survey, 2XMM~J123103.2+110648 (hereinafter J1231+1106 for brevity).  The X-ray 
properties of this source are exceptional in at least two respects.  Its X-ray 
light curve shows not only strong variability (factor of 3 or more within 1000 
s) but, in at least one of its three observations, possibly also 
quasi-periodic modulations with a period of $\sim 15,000$ s.  More remarkable 
still is its X-ray spectrum.  The spectrum of J1231+1106 completely lacks 
emission at energies greater than $\sim 2$ keV and can be described entirely 
by a soft thermal component.  No significant intrinsic X-ray absorption is 
detected.  Such an extreme soft excess is unprecedented among AGNs.  Modeling 
the soft excess with a multicolor disk blackbody yields an inner disk 
temperature of $kT = 0.16-0.21$ keV.  Moreover, the soft X-ray band shows 
spectral variability consistent with Comptonization.  Both the strength of the 
soft excess and the possible evidence for Comptonization suggest that the 
source has a high Eddington ratio (\gax\ 0.3).  In combination with the 
observed luminosity, the BH may be no more massive than $\sim 10^5$ \solmass, 
one of the lowest ever identified in an AGN.  

Interesting though they may be, the above results strongly hinge on the 
correct identification of J1231+1106 as an extragalactic source.  Terashima et 
al. (2012) argue that it is unlikely to be a Galactic source.  Based on its 
positional coincidence, they tentatively associate J1231+1106 with the optical 
counterpart SDSS~J123103.24+110648.5, a small ($r$-band isophotal diameter 
$\sim 2$\asec), faint ($r = 20.04\pm0.04$ AB model mag), but rather red ($g - 
r = 0.83\pm0.04$ mag) galaxy with an estimated photometric redshift of $z = 
0.13\pm0.05$ according to the Seventh Data Release of the Sloan Digital Sky 
Survey (SDSS; Abazajian et al. 2009).  Assuming a distance of $D_L$ = 610 Mpc 
(based on a cosmology of $H_0 = 70$~\kms~Mpc$^{-1}$, $\Omega_{m} = 0.3$, and 
$\Omega_{\Lambda} = 0.7$), the intrinsic $0.5-2$ keV luminosity is 
$L_{\rm X} = (1.6-3.8) \times 10^{42}$ \lum, where the range reflects the 
extremes of the observed variations.  If we assign all of the optical emission 
to starlight, the galaxy has a maximum absolute magnitude of $M_g = -18.1$, 
characteristic of dwarf galaxies and within the range of other AGNs with 
low-mass BHs (Greene \& Ho 2004, 2007c).  However, even if the X-ray source is 
associated with the faint galaxy, it is unclear whether the emission comes 
from the galaxy nucleus or, instead, from an exceptionally powerful 
off-nuclear (i.e. ultraluminous or hyperluminous) X-ray source.

\vskip 0.3cm
\begin{figure*}[t]
\centerline{\psfig{file=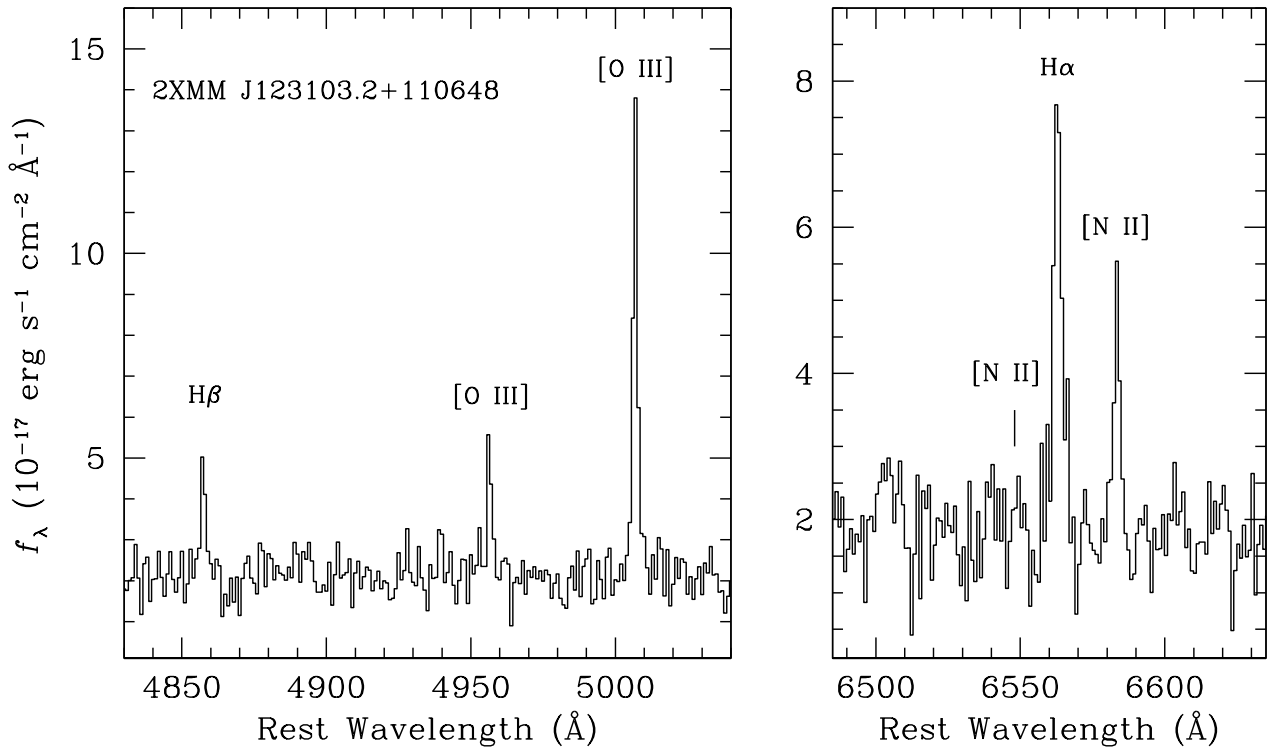,width=17.5cm,angle=270}}
\figcaption[fig1.ps]{
Magellan spectrum of J1231+1106, with major emission lines labeled.
The blue portion of the spectrum has been scaled to match the red portion,
and a redshift of $z = 0.1187$ has been removed.
\label{fig1}}
\end{figure*}
\vskip 0.3cm

Using a new optical spectrum of J1231+1106, we present an accurate measurement 
of the redshift of this source, confirm that it is an AGN based on the 
excitation of its nebular lines, although no broad permitted lines are 
visible, and use the velocity dispersion of the narrow emission lines to 
constrain the mass and Eddington ratio of the central BH.

\section{Spectroscopic Observations}

We observed J1231+1106 using the Inamori Magellan Areal Camera and 
Spectrograph (IMACS) on the Baade 6.5 meter telescope at Las Campanas 
Observatory.  Two spectra were taken in long-slit mode with the 600 
lines~mm$^{-1}$ grating, one on 2012 May 14 UT blazed for blue coverage 
(3650--6600 \AA) using a 0\farcs7 slit and another on 2012 July 14 UT blazed 
for red coverage (5700--8900 \AA) using a 1\farcs2 slit.  Total integration 
times were 3000~s and 1800~s, respectively, split into multiple exposures. The 
sky conditions were clear and the seeing average ($\sim$0\farcs65--1\farcs15). 
Both observations were taken at moderately high airmasses (1.4--1.6), but the 
slit was oriented along the parallactic angle to minimize slit losses due to 
differential atmospheric refraction.  The instrumental resolution, estimated 
from the full width at half maximum (FWHM) of the night sky lines, is 1.2 \AA\ 
for the blue setting and 3.5 \AA\ for the red setting, which corresponds to a 
velocity dispersion of $\sigma_{\rm inst} = 31.8$ \kms\ for \oiii\ \lamb 5007 
and $\sigma_{\rm inst} = 68.1$ \kms\ for \ha.

Data reduction was performed using tasks in IRAF\footnote{IRAF is distributed 
by the National Optical Astronomy Observatory, which is operated by the 
Association of Universities for Research in Astronomy (AURA) under cooperative 
agreement with the National Science Foundation.}, following basic procedures 
outlined in Ho \& Kim (2009).  In brief, the two-dimensional data were 
corrected for bias and flat-fielded using a series of external quartz lamp 
images.  Cosmic rays were removed by combining multiple exposures using 
``crreject'' in {\tt imcombine}.  The extracted one-dimensional spectra 
were wavelength calibrated using He+Ne+Ar arc lamp spectra and then corrected 
for telluric absorption and flux calibrated using spectra of the 
spectrophotometric standard stars Feige 56 and LTT 4816.

\section{Optical Properties}

Figure~1 shows the final optical spectrum of J1231+1106.  The flux density 
scale of the red spectrum matches closely with that expected from the SDSS 
photometry, confirming that our absolute spectrophotometric calibration is 
reasonably accurate.  We measure $f_\lambda$(6230~\AA) = $(2.4\pm0.3) \times 
10^{-17}$ \flamb; this corresponds to $r = 20.17\pm0.13$ AB mag, to be 
compared to $r = 20.04\pm0.04$ AB mag from SDSS photometry.  The blue and red 
portions of the spectrum agree reasonably well in shape in their region of 
overlap ($\sim 5800-6500$ \AA), but the absolute flux scale on the blue side 
is 41\% lower, presumably because of the narrower slit used in the blue 
setting.  We have accordingly scaled the blue spectrum by a factor of 1.41.

We detect a weak continuum and sharp, narrow emission lines from \hb, \oiii\ 
\lamb\lamb4959, 5007, \ha, and \nii\ \lamb6583.  \nii\ \lamb6548 is absent, 
but its upper limit is consistent with $\sim 1/3$ of \nii\ \lamb6583.  No 
broad component is seen in \ha\ or 

\psfig{file=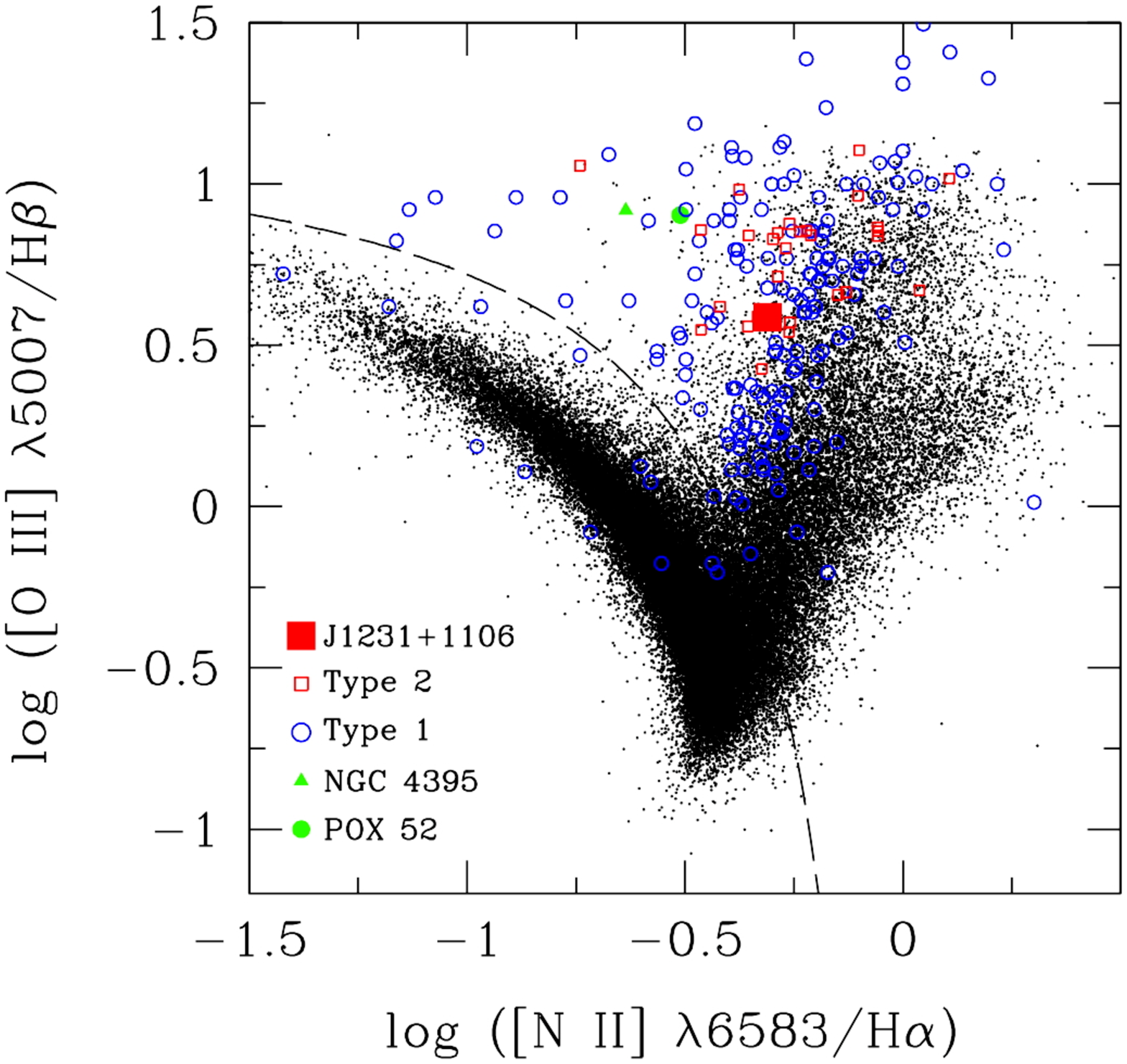,width=8.75cm,angle=0}
\vskip 0.3cm
\figcaption[fig2.ps]{
Line-ratio diagnostic diagram of \oiii\ \lamb5007/\hb\ vs.  \nii\
\lamb6583/\ha.  The small black points are SDSS measurements from Kauffmann et
al. (2003) with signal-to-noise ratios larger than 6.  J1231+1106 is marked
with a large, filled red square.  Small open red squares are low-mass type 2
Seyferts from Barth et al. (2008), while the type 1 sample of Greene \& Ho
(2007a) is shown as open blue circles.  NGC~4395 and POX~52 (Barth et al.
2004) are plotted as a filled green triangle and circle, respectively.  The
dashed line represents the empirical boundary proposed by Kewley et al. (2006)
to separate star-forming galaxies (left) from AGNs (right).
\label{fig2}}
\vskip 0.3cm

\noindent
\hb.  The presence of moderately strong 
\oiii\ \lamb5007 emission compared to \hb\ (3.86$\pm$0.04) and significant 
\nii\ \lamb 6583 relative to \ha\ (0.49$\pm$0.05) qualifies the source as an 
AGN.  In the diagnostic diagram of \oiii\ \lamb5007/\hb\ vs.  \nii\ 
\lamb6583/\ha\ (Baldwin et al. 1981; Figure~2), J1231+1106 lies comfortably 
within the locus of high-excitation AGNs (Seyfert nuclei), but, not unlike 
other low-mass AGNs, it is somewhat displaced toward lower \nii\ 
\lamb6583/\ha, presumably because of the lower metallicity expected for the 
low-luminosity, low-mass host galaxy (Ludwig et al. 2012).

Our spectra barely resolve the lines in the blue but not at all 
in the red.  The intrinsic width of \oiii\ \lamb 5007, the strongest line, is 
FWHM = 78.8$\pm$6 \kms, or $\sigma = 33.5\pm2.6$ \kms\ for a Gaussian profile.
This is among the narrowest emission line widths ever measured for AGNs (Barth 
et al. 2008; Xiao et al. 2011).  The centroid of  \oiii\ \lamb5007 yields an 
accurate measurement of the redshift, $z = 0.11871\pm0.000002$; \ha\ gives a 
consistent value of $z = 0.1176 \pm 0.00002$, but we adopt that based on 
\oiii\ because of its higher precision.  For the same set of cosmological 
parameters adopted by Terashima et al. (2012), the revised distance for 
J1231+1106 is $D_L = 553$ Mpc.  

Assuming an intrinsic ratio of \ha/\hb\ = 3.1 for AGNs (Halpern \& Steiner 
1983) and the extinction curve of Cardelli et al. (1989), the observed Balmer 
decrement of \ha/\hb\ = 3.55 corresponds to an internal extinction of $A_V = 
0.425$ mag.  Galactic extinction contributes an additional $A_V = 0.092$ mag 
(Schlafly \& Finkbeiner 2011).  Correcting for a total line-of-sight 
extinction of  $A_V = 0.517$ mag, J1231+1106 has intrinsic line luminosities of 
$L_{\rm [O~III]} = 1.66 \times 10^{40}$ \lum\ and $L_{\rm H\alpha} = 1.32 
\times 10^{40}$ \lum.

To gain additional insight into the nature of the host of J1231+1106, we also 
performed a simple analysis of the IMACS acquisition image recorded during 
the second observing run.  The 80~s $V$-band exposure is shallow, and the 
seeing was poor (FWHM $\approx$ 1\farcs15), but the pixel scale of the $f$/4 
camera (0\farcs11) does afford enough resolution to provide some rudimentary 
constraints on the photometric structure of the galaxy.  We used the code
GALFIT (Peng et al. 2010) to construct a simple two-dimensional model, which 
consists of a single S\'ersic (1968) function convolved with a point-spread 
function derived from a nearby star on the same CCD chip as the galaxy.
The best-fit model yields a S\'ersic index of $n \approx 0.7$, close to an 
exponential profile ($n = 1$), and an effective radius of $R_e$ = 0\farcs32 
$\simeq 0.7$ kpc.

\section{Physical Implications}

We have conclusively established that the highly variable, unusually soft X-ray
source J1231+1106 is of extragalactic origin at $z = 0.11871\pm0.000002$.  The 
optical spectrum is dominated by narrow emission lines excited by a nonstellar 
mechanism, presumably photoionization by the same source producing the X-rays.
This confirms that J1231+1106 is indeed an AGN.

No broad \ha\ or \hb\ is seen in the spectrum, making this a type 2 AGN not 
dissimilar from those reported by Barth et al. (2008).  A key distinction, 
however, is that while the Barth et al. (2008) sources are highly absorbed in 
the X-rays (Thornton et al. 2009), J1231+1106 is not.  Its {\it XMM-Newton}\ 
spectrum shows no evidence of any significant intrinsic X-ray absorption.  
Comparing the X-rays with the strength of the \oiii\ emission further supports 
this, as illustrated in Figure~3.  For reference we show the best-fitting 
relation between $L_{\rm X}$ and $L_{\rm [O~III]}$ for luminous, unabsorbed 
AGNs (Panessa et al. 2006), along with low-mass type 1 AGNs (Greene \& Ho 
2007a; Desroches et al. 2009; Dong et al. 2012), low-mass type 2 AGNs 
(Thornton et al. 2009), and the two low-mass type 1 archetypes NGC~4395 
(Panessa et al. 2006) and POX~52 (Thornton et al. 2008).  For consistency with 
the J1231+1106 measurement, all literature X-ray luminosities have been 
adjusted to the 0.5--2 keV band using published photon indices whenever 
available or assuming $\Gamma = 1.8$ when not.  Unexpectedly, the ratio of 
X-ray to \oiii\ luminosity in J1231+1106---optically a type 2 source---lies 
securely within the range occupied by unabsorbed type 1 AGNs, both of high and 
low mass.  By contrast, the four low-mass type 2 systems studied by Thornton 
et al. (2009) all sit substantially below the bulk of the type 1 AGNs in 
Figure~3, consistent with the X-rays being significantly absorbed in these 
systems.

Taken at face value, the lack of X-ray absorption in an ostensibly type 2 
Seyfert violates the conventional geometric unification picture for AGNs 
(Antonucci 1993).  In this sense J1231+1106 joins a small number of AGNs that
seem to intrinsically lack a broad-line region (e.g., Ghosh et al. 2007; 
Brightman \& Nandra 2011; Tran et al. 2011; Matt et al. 
2012).  Could the apparent absence of broad lines be an artifact?
Probably not. If the BH mass is as low as $\sim 10^5$ \solmass\ (see below), we 
expect any broad \ha, if present, to be inherently weak and challenging to 
detect.  Our chances of seeing it, however, could have been compromised by the 
relatively low signal-to-noise ratio of the \ha\ portion of the spectrum.  To 
estimate an upper limit to the amount of broad \ha\ allowed by the data, we 
assume that the hypothetical feature has the median line width of the sample 
of low-mass type 1 sources from Greene \& Ho (2007c).  Given \mbh\ = $10^5$ 
\solmass\ and FWHM = 900 \kms, the \ha-based virial mass estimator of Greene 
\& Ho (2005b; as updated in Xiao et al. 2011) predicts $L_{\rm H\alpha} 
\approx 5 \times 10^{40}$ \lum, roughly 4 times the observed luminosity of 
narrow \ha.  Simple experimentation with the data shows 

\psfig{file=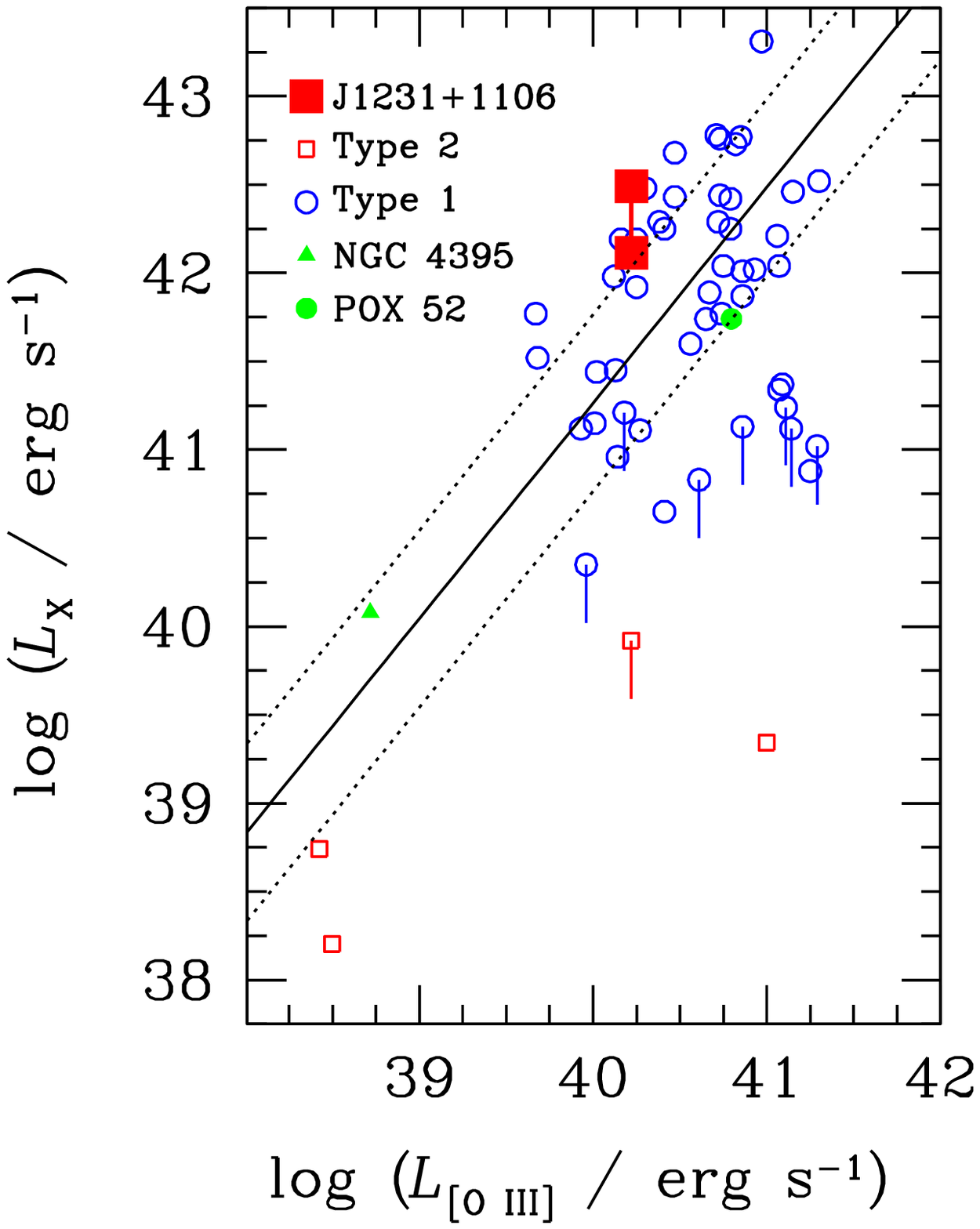,width=8.75cm,angle=0}
\figcaption[fig3.ps]{
Correlation between absorption-corrected soft X-ray luminosity in the
0.5--2 keV band and \oiii\ \lamb5007 luminosity for J1231+1106 (two
large, filled red squares, representing the highest and lowest flux values;
Terashima et al. 2012), low-mass type 2 AGNs (open red squares; Thornton et
al. 2009); low-mass type 1 AGNs (open blue circles; Greene \& Ho 2007a;
Desroches et al. 2009; Dong et al. 2012), NGC~4395 (filled green triangle;
Panessa et al. 2006, adjusted to a distance of 4.3 Mpc),
and POX~52 (solid green circle; Thornton et al. 2008).  The solid line is the
best-fitting relation from Panessa et al. (2006), with the 1~$\sigma$ scatter
denoted by the two dotted lines; their X-ray luminosities were transformed
from the 2--10 keV band to the 0.5--2 keV band assuming $\Gamma = 1.8$.
\label{fig3}}
\vskip 0.3cm

\noindent
that it is difficult 
to hide a broad \ha\ line of this strength, leading us to conclude that 
1231+1106 is genuinely a type 2 source.

As with the type 2 sources studied by Barth et al. (2008) and Thornton et al. 
(2009), we anticipate that the BH mass in J1231+1106 must be very low.  
Several aspects of the optical data presented in this study lend themselves to 
this interpretation.  First, the host galaxy is physically small ($R_e \approx 
0.7$ kpc), disk-dominated (S\'ersic index $n \approx 1$), low-luminosity ($M_g 
= -17.9$ mag, corrected to the revised distance and for a Galactic extinction 
of $A_g = 0.111$ mag), and, judging by the narrowness of its emission lines 
($\sigma = 33.5$ \kms), presumably low-mass.  For comparison, J1231+1106 is at 
least 2 magnitudes fainter than typical field galaxies at $z=0.1$, which have 
a characteristic luminosity of $M^*_g = -20.1$ mag (Blanton et al. 2003), and 
over a magnitude fainter than the hosts of most previously studied low-mass 
type 1 ($\langle M_g \rangle = -19.3$ mag; Greene \& Ho 2007c) or type 2 
($\langle M_g \rangle = -19.0$ mag; Barth et al. 2008) AGNs.   The location of 
J1231+1106 on the \oiii\ \lamb5007/\hb\ vs. \nii\ \lamb6583/\ha\ diagnostic 
diagram additionally hints that the gas-phase metallicity may be sub-solar.  
These gross properties remind us of bright dwarf galaxies or very late-type 
spirals.  They also share strong morphological similarities with the host 
galaxies of low-mass type 1 AGNs such as NGC~4395 and many of the Greene \& Ho 
(2004, 2007c) objects (e.g., Jiang et al. 2011).  Curiously, 
however, J1231+1106 has exceptionally red optical colors ($u-g=1.53$ mag; 
$g-r=0.83$ mag) for its luminosity and presumed late morphological type.  
Late-type spirals have $g-r$ \lax\ 0.5 mag, and even S0 galaxies are seldom 
redder than $g-r \approx 0.7$ mag (Fukugita et al. 1995).  If the host of 
J1231+1106 is, in fact, disk-dominated, it must be mildly reddened by dust.

The current set of optical observations places an independent constraint on the
BH mass and Eddington ratio of J1231+1106.  The lack of detectable broad
permitted lines prevents us from estimating the mass of the central BH using
conventional methods developed for type 1 AGNs.  However, we can offer an
educated guess using the width of the narrow lines.  The velocity dispersion
of the narrow-line region gas in AGNs generally traces the gravitational
potential of the stars in the host, such that, to first order, $\sigma_g
\approx \sigma_*$.  This rough approximation holds
for a wide range of AGN types and activity levels (Greene \& Ho 2005a; Ho
2009), including low-mass systems (Barth et al. 2008; Xiao et al. 2011).
Moreover, the $M_{\rm BH}-\sigma_*$ relation extends across a wide range of
masses, including $M_{\rm BH}$ \lax\ $10^6$ \solmass\ (Barth et al. 2005;
Greene \& Ho 2006; Xiao et al. 2011).  Adopting $\sigma_* \simeq \sigma_g =
33.5$ \kms, the $M_{\rm BH}-\sigma_*$ of G\"ultekin et al. (2009) for inactive
galaxies predicts $M_{\rm BH} = 6.8 \times 10^4$ \solmass.  Active galaxies,
especially toward lower masses, may define a shallower $M_{\rm BH}-\sigma_*$
relation.  The fit of Xiao et al. (2011) yields a slightly larger mass of
$M_{\rm BH} = 1.3 \times 10^5$ \solmass.  Although the $M_{\rm BH}-\sigma_*$
relation at these low masses is highly uncertain (Xiao et al. 2011 estimate a
formal intrinsic scatter of 0.46 dex), we can safely conclude that the BH
powering J1231+1106 may be as low as $M_{\rm BH} \approx 10^5$ \solmass.  Such
a value would be in accord with the general characteristics of the host
(discussed above), as well as the arguments based on the X-ray properties
mentioned in the Introduction.

Following the methodology of Greene \& Ho (2007b), the extinction-corrected
\ha\ luminosity of $L_{\rm H\alpha} = 1.32 \times 10^{40}$ \lum\ (Section 3)
implies $L_{\rm bol} = 2.34\times 10^{44} (L_{\rm H\alpha}/10^{42})^{0.86} =
5.66 \times 10^{42}$ \lum, or $L_{\rm bol}/L_{\rm Edd} = 0.45$ for $M_{\rm BH}
= 10^5$ \solmass.  If we adopt the lower value of $M_{\rm BH} = 6.8\times10^4$
\solmass, $L_{\rm bol}/L_{\rm Edd} = 0.66$.  Either of these two values is
in excellent agreement with the expectation that the Eddington ratio should
exceed $\sim$0.3 to account for the observed Comptonization of the X-ray
spectrum (Terashima et al. 2012).

Lastly, we computed a few simple photoionization models using the code CLOUDY 
(Ferland et al. 1998) to test whether the unusual X-ray spectrum of J1231+1106 
can be responsible for the excitation of its optical lines.  We assume that 
the ionizing continuum can be described by a blackbody with $T=1.7 \times 
10^6$ K, corresponding to a 0.15 keV blackbody that can approximate the 
X-ray spectrum.  We vary the ionization parameter in the range
$U = 10^{-4.5}- 10^{-1.5}$ and the hydrogen density $n_{\rm H} = 10^2 - 10^6$
\cc, values typical of the narrow-line regions of most AGNs (e.g., Nagao et al.
2001; Kim et al. 2006).  For simplicity we assume solar metallicity,
plane-parallel geometry, and constant hydrogen density and neglect the effects
of dust.  These calculations confirm that the observed X-ray spectral
energy distribution of J1231+1106 can, indeed, reproduce line ratios that
span the locus of AGNs in standard diagnostic diagrams such as that shown
in Figure~2.  Given the simplicity of the models, we did not attempt to
fine-tune them to exactly match J1231+1106.

\clearpage
\acknowledgements
This work is supported by the Carnegie Institution for Science (LCH), a
KASI-Carnegie Fellowship (MK), and Grant-in-Aid for Scientific Research 
20740109 (YT) from the Ministry of Education, Culture, Sports, Science, and 
Technology of Japan.


\end{document}